\documentclass[9pt,twocolumn,twoside]{osajnl}

\journal{ol} 

\setboolean{shortarticle}{true}

\usepackage{lineno}

\title{Hyper-spectral imaging through a multi-mode fiber}

\author[1]{Al\.{ı}m Yolalmaz}
\author[2]{Emre Y\"{u}ce}

\affil[1]{Department of Imaging Physics, Delft University of Technology, Lorentzweg 1, 2628 CJ Delft, The Netherlands}
\affil[2]{Programmable Photonics Group, Department of Physics, Middle East Technical University, 06800 Ankara, Turkey}

\affil[*]{Corresponding author: A.Yolalmaz@tudelft.nl}

\begin{abstract}

Multi-mode fibers provide an increased amount of data transfer rates given a large number of transmission modes. Unfortunately, the increased number of modes in a multi-mode fiber hinders the accurate transfer of information due to interference of these modes which results in a random speckle pattern. The complexity of the system impedes the analytical expression of the system thereby the information is lost. However, deep learning algorithms can be used to recover the information efficiently. In this study, we utilize deep learning architecture to reconstruct input colored images from the output speckle patterns at telecommunication wavelength (C-band). Our model successfully identifies hyper-spectral speckle patterns at twenty-six separate wavelengths and twenty-six distinct letters. Remarkably, we can reconstruct the complete input images only by analyzing a small portion of the output speckle pattern. Thereby, we manage to decrease the computational load without sacrificing the accuracy of the classification. We believe that this study will show a transformative impact in many fields: biomedical imaging, communication, sensing, and photonic computing.

\end{abstract}

\begin{document}

\maketitle

\section{Introduction}

Optical fiber has been the key component in many applications that include communication\cite{Uden2014, Richardson2013}, holographic optical manipulation\cite{Leite2017, Bianchi2012}, lensless focusing\cite{Amitonova2016}, optical coherence tomography imaging\cite{Huang1991}, fiber-optic fluorescence imaging\cite{Flusberg2005}, chemical sensing\cite{Yolalmaz2019}, and monitoring biological tissues\cite{Wu2009, Ohayon2018}. Utilization of optical fibers, especially in imaging, is of utmost importance, which provides superior resolution and a small footprint for extreme visualization purpose\cite{Wu2009, Ohayon2018, Andresen2013, Amitonova2016}. Compared to a single-mode fiber, a multi-mode fiber (MMF) delivers much more information due to a larger core diameter resulting in an increased number of modes. When the number of modes increases, the information amount propagating through the MMFs, such as the number of pixels in a scene, increases. However, with coherent light, which is required to carry controlled light propagation, the wavefront of light is scrambled due to inter-modal mixing through the MMF because each of these individual modes propagates at a slightly different velocity\cite{Yariv1976, Mitschke2010}. Therefore, the image at the fiber output appears as a random array of bright and dark spots referred to as a complex spatiotemporal speckle pattern\cite{Goodman2000, Goodman2007}. This effect completely scrambles the input image.

There are many innovative techniques for imaging with an MMF. One of them is based on the propagation of images through the MMF with principles of digital holography\cite{Cizmar2012}. Also, digital phase conjugation enables to form the images at the end of the MMF\cite{Papadopoulos2013}. By measuring the experimental transmission matrix of an MMF, image propagation through the MMF is performed in a controlled mean\cite{Ploeschner2015, Choi2012, Loterie2015, Fan2021}. Unfortunately, the aforementioned techniques require complex experimental setups with a reference path, high-resolution speckle patterns, or a short fiber length. With the advancement in computer processing power offered by graphical processing units and progress in artificial intelligence\cite{LeCun2015}, image transfer through the MMF is achieved with neural networks in a few seconds\cite{Takagi2017, Borhani2018, Rahmani_2018, Fan2019a, Kueruem2019, Wang2018}. The neural network encoders infer the images from the speckle patterns without any need for a prior mathematical model of the fiber. However, the resolution of a speckle pattern strongly affects the fidelity of a reconstructed image and should be higher or equal to the resolution of an input image\cite{Kueruem2019}. Further improvement of fidelity is of paramount importance as it can provide a more precise signal in optical communications and sharper images in medical diagnosis. Weak similarities among the speckle patterns make identification and discrimination of the input images easier, but the previous works do not address the similarities among the speckle patterns, which strongly bears importance to accurately classifying and reconstructing images with the speckles. Also, the diversity of information that is encoded into the speckles is not taken into account to identify the input images transferred through the MMFs.

In this manuscript, we present hyperspectral imaging through an MMF using a spatial light monitor (SLM) to form the images. A tunable laser C-band is used to alter the wavelength of the input image. To reveal information hindered by inter-modal mixing through the MMF, we developed neural networks to reconstruct hyper-spectral images and classify corresponding speckle patterns at telecommunication wavelength. Instead of training the neural networks with full resolution of the speckle patterns, which increases the number of models' weights and training duration of models, we develop our models using 1/400 of the speckle images. Moreover, we aim to enrich the diversity of information propagating through the MMF using 16900 different letter images. We inspect the similarity among the speckle patterns within the data set to evaluate the complexity of image classification and reconstruction in detail. Our neural networks reconstruct and classify color images in two seconds. Compared to the use of existing neural networks such as UNet, Resnet, and VGG-Net\cite{Rahmani_2018}, our neural networks are specifically designed for imaging through the MMF.

\section{Methods}

\subsection{Experimental setup}

\begin{figure}[!htb]
    \centering
        \includegraphics[width=82 mm]{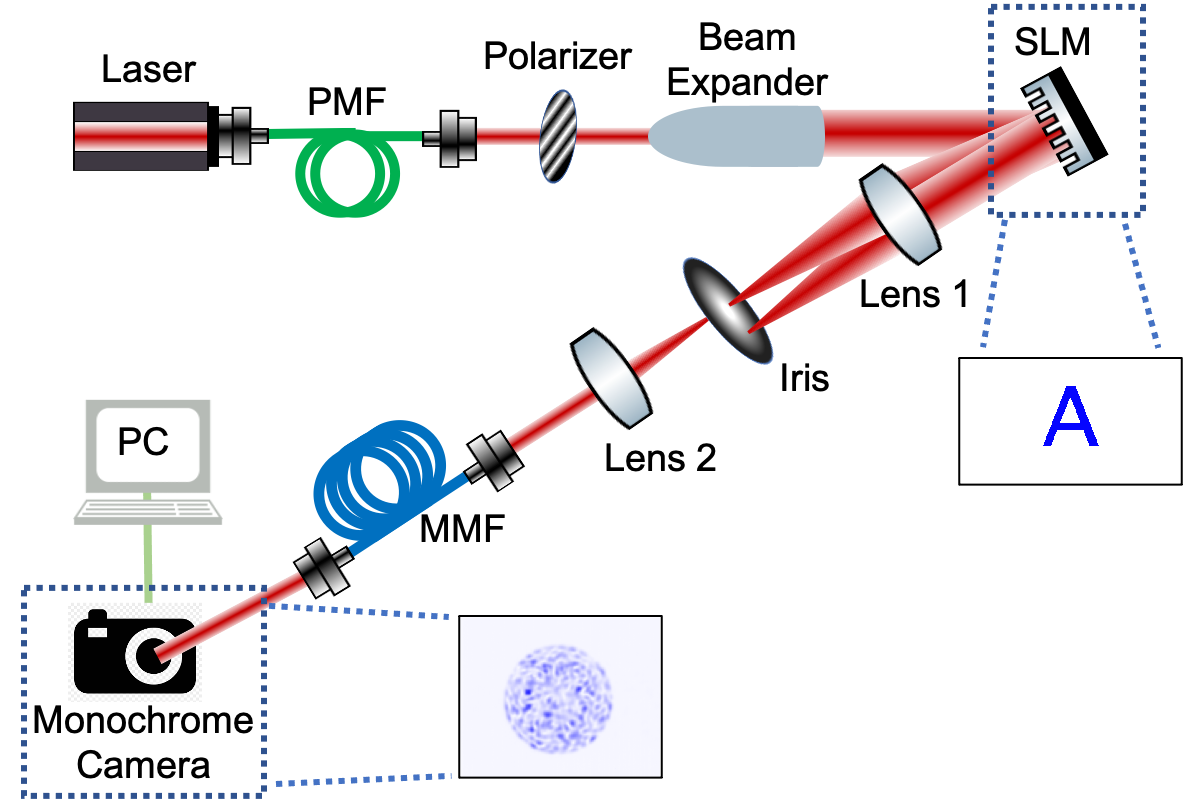}
        \caption{The scheme of our experimental setup for imaging through a multi-mode fiber. The setup consists of a laser, a polarization-maintaining fiber (PMF), an achromatic half-wave wave plate polarizer, a beam expander, a spatial light modulator (SLM), two convex lenses, an iris, a multi-mode fiber (MMF), a monochrome camera, and a personal computer (PC).}
    \label{setup}
\end{figure}

The experimental setup for delivering an image through an MMF is presented in Fig. \ref{setup}. The continuous laser beam from a fiber-coupled laser source (Santec Inc., Wavelength selectable laser, Part no: WSL-100) with a 1 pm spectral resolution is first coupled into a polarization-maintaining fiber. Next, the beam is coupled to free space, and then the polarization direction of the laser is rotated by an achromatic half-wave plate polarizer. Then, the size of the laser beam is expanded by using a beam expander, which provides 15 times enlargement in size of the laser beam. Later, the beam is reflected from a spatial light modulator (SLM, 1920x1080, Holoeye Inc., Pluto-2.1-Telco-013), enabling 8-bit pixel-wise phase modulation of light with a frame rate of 60 Hz. Phase modulation by the SLM is presented with a phase plate, a gray-scale image whose pixel value ranges between 0 and 250. A 4f configuration (Lens 1 and Lens 2) is used to compress the beam and direct it to the fiber collimator. The MMF has a step-index core with a core diameter of 105 $\mu$m, a NA of 0.22, and around 1096 propagating optical modes at 1550 nm. The MMF does not preserve the polarization direction of the light, which is chaotic in the MMF. The beam travels through the 20 m MMF and is then coupled out from the fiber. Next, the beam is delivered to a monochrome camera (Allied Vision Inc., Part no: Goldeye G-008) which generates an 8-bit gray-scale image with a resolution of 320-by-256 and a pixel size of 30 $\mu$m.

In our study, we use the first diffraction order of the beam reflected from the SLM. The reason is that the 0th order cannot be fully controlled and presents an unmodulated signal on the camera. Working on the first diffraction order allows us to efficiently control light intensity on the monochrome camera without considering interference with unmodulated light on the zeroth-order. For the first diffraction order generation, we modulate the laser with the SLM, which behaves as a blazed grating and yields diffraction orders. For image transmission through the MMF, we sum a phase pattern of an image and the phase pattern of the blazed grating to transmit the image and simultaneously obtain the first diffraction order of the laser beam. If a pixel of the summed phase pattern shows phase modulation beyond 2$\pi$, we use modular arithmetic to bring phase shift in the pixel between 0 and 2$\pi$. An example of a phase pattern, which is for an image of letter A displayed on the SLM without phase of blazed grating, is presented in Fig. \ref{setup}. An example speckle pattern on the monochrome camera (see Fig. \ref{setup}) proves that many of the supported modes in the MMF are excited at the fiber input. We also observe that the speckles are randomly and uniformly distributed over the fiber end facet.

\subsection{Data set}

We generate English alphabet images from A to Z for modulation of the laser beam by the SLM. The resolution of the letter images is the same as the resolution of the SLM, 1920-by-1080 pixels. The phase shift between the background of letters and letters changes between 0.14$\pi$ and 3.42$\pi$ with a step of 0.14$\pi$ at 1550 nm. The images of letters have twenty-five different phase levels, so different phase shifts generate different speckle patterns in the monochrome camera. The resolution of the speckle patterns equals 200-by-200 after cropping the monochrome camera images to discard the unilluminated part of the camera. The wavelength of the tunable laser source is altered from 1538 nm to 1563 nm with a 1 nm wavelength step to monitor the speckle patterns under different wavelengths of the laser. At the end of the data generation process, we obtain 16900 input images and output speckle patterns. The overall data collection time is 10 hours, and the monochrome camera's data acquisition rate and the SLM's refresh rate strongly affect the data collection time.

\begin{figure}[!htb]
    \centering
    \includegraphics[width=80 mm]{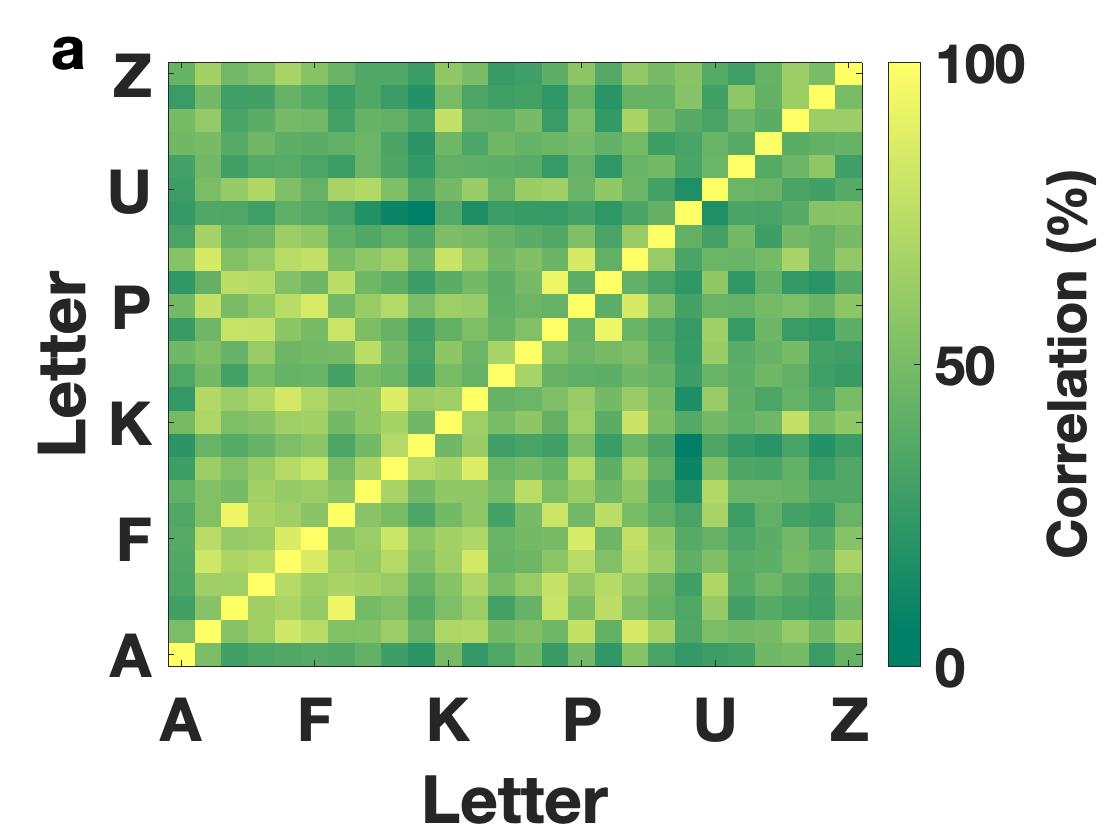}\\
        \includegraphics[width=80 mm]{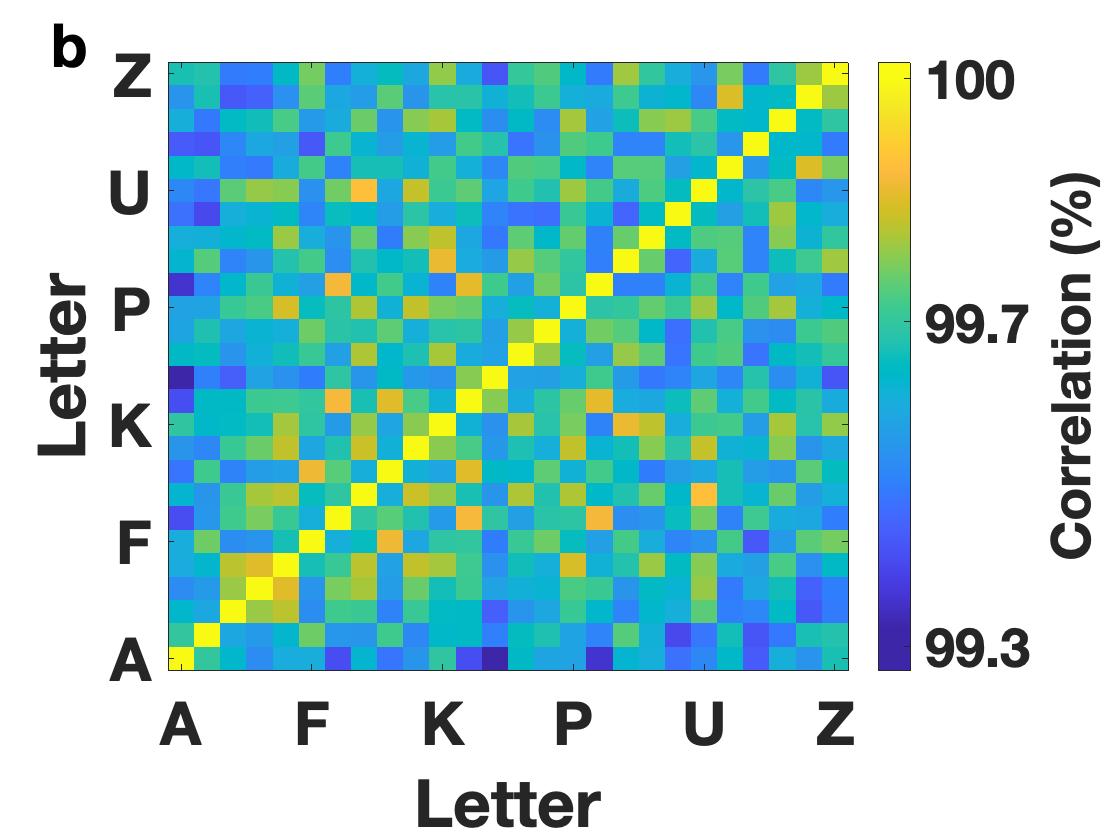}\\
        \includegraphics[width=80 mm]{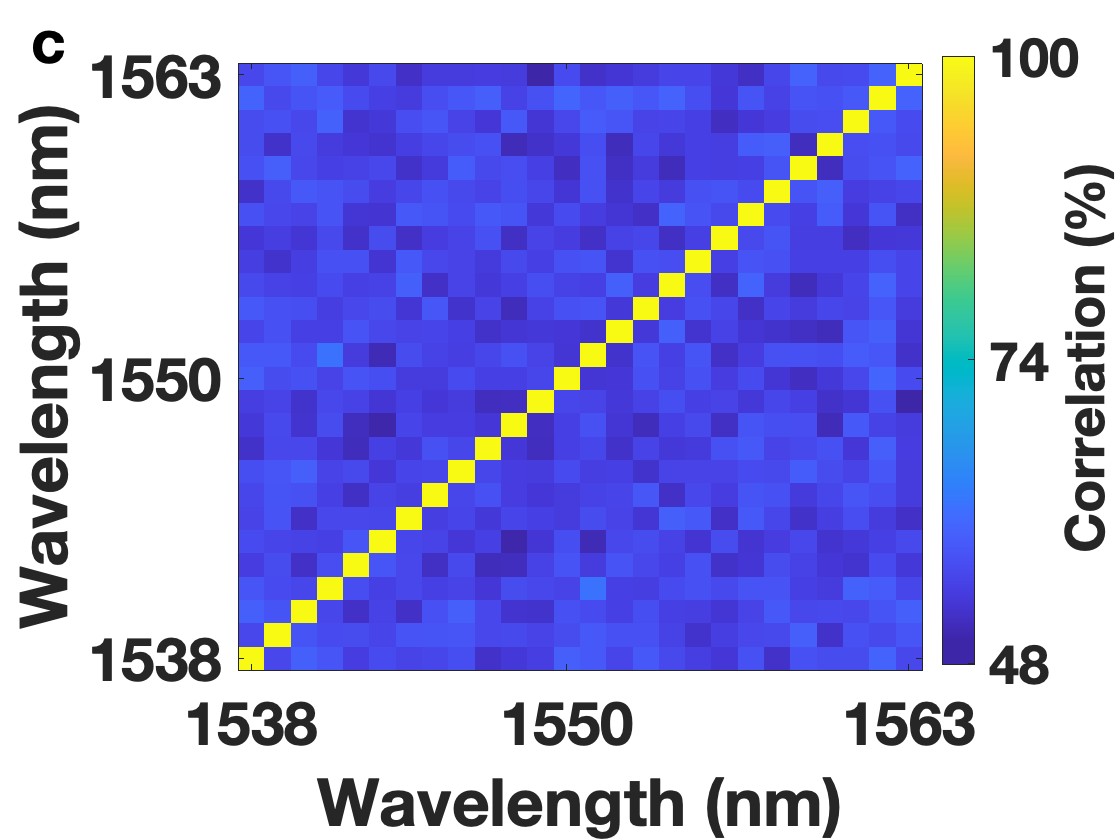}
    \caption{(a) Correlation map of twenty-six English letter images, (b) Correlation map of corresponding speckle patterns, (c) Correlation map of speckle patterns for letter A when the wavelength of the laser changes between 1538 nm and 1563 nm.}
    \label{maps}
\end{figure}

Through this study, the Pearson correlation coefficient performs to evaluate similarity among images of different letters. The same metric expresses similarity among different speckle patterns for corresponding letter images. Also, the Pearson correlation coefficient is utilized to calculate the image reconstruction accuracy. For this purpose, the Pearson correlation coefficient is computed with true letter images and predicted letter images via neural networks. We name the Pearson correlation coefficient as the correlation for simplicity in this study, and it is presented in percent.

In this study, we aim to increase the diversity of information propagating through the MMF to enhance the amount of information delivered from the SLM to the camera. For this purpose, we share a correlation map for different letter images in Fig. \ref{maps}a. Image correlation varies between 5\% and 94\% in Fig. \ref{maps}a. However, the correlation of the speckle patterns at 1538 nm for different letter images does not sharply change and varies between 99.3\% and 99.9\% (Fig. \ref{maps}b). Compared to strong changes in letter image correlations, this low variation between the speckle patterns is due to the blaze grating pattern, which forms the diffraction orders. Also, the monochrome camera we operate has an intensity threshold, and the light intensity is not detected below that value. Thus, this issue decreases the accuracy of light detection, leading to a high correlation between the speckle patterns for different letter images. We also share a correlation map for letter image A when the incident laser wavelength varies between 1538 nm and 1563 nm in Fig. \ref{maps}c. When the laser wavelength changes, the correlation of speckle patterns at two different wavelengths is strongly affected and varies between 49.9\% and 61.5\% (correlation of an image with itself is 100 \% as expected in Fig. \ref{maps}).

\subsection{Deep neural networks}

In this work, we use convolutional neural network (CNN) architecture to reconstruct the letter images from the speckle patterns and classify the speckle patterns in terms of letter and wavelength (see Supplementary material for neural networks). The CNN layers have remarkable properties: sharing layers' weights, the existence of sparse filters, and convolution operation. Thus, CNN layers decrease the computation cost of deep learning models. Our models also constitute dense layers to unravel hidden information between the speckle patterns and the letter images. To diminish the computational load of the models, we take a 10-by-10 portion of the speckle patterns to train the models and correlate the speckle patterns with letter images of 64-by-36 resolution after max-pooling the original letter image resolution 1920-by-1080. The models' weights are statically decided by considering the speckle patterns and the letter images. Tuning the weights is repeated for a fixed number of epochs, 200 for classification, and 400 for the reconstruction of letter images while ensuring convergence of categorical cross-entropy cost function to a minimum value. We train the models with 80\% of the data and validate their performance with 20\% of the data. We openly share our deep learning models as supplementary material to contribute to the community to boost image classification and reconstruction from speckle patterns.

\section{Results and discussion}

\begin{figure}[!htb]
    \centering
    \includegraphics[width=84 mm]{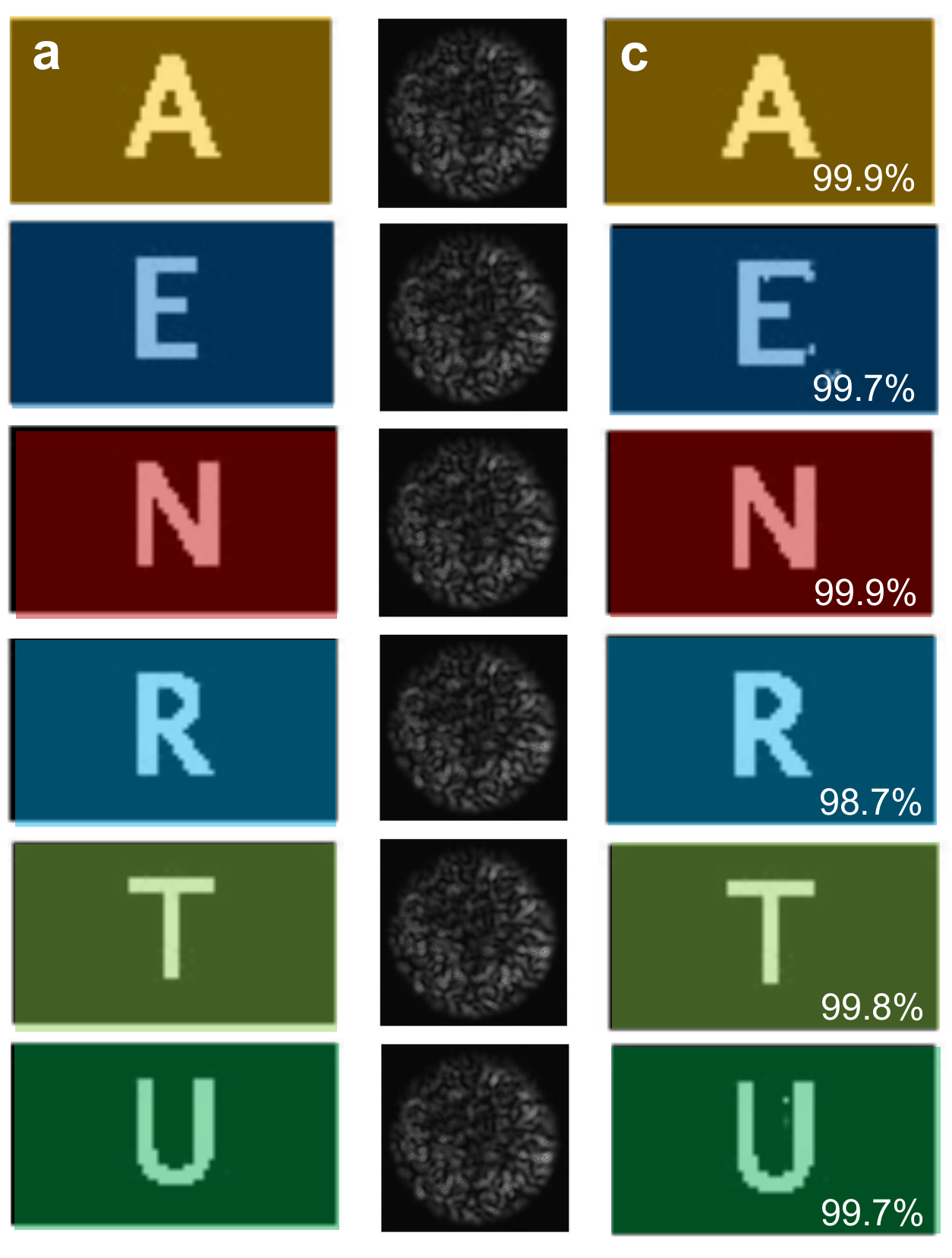}
        
    \caption{(a) True letter images which propagate through the MMF. (b) Speckle patterns at the monochrome camera for corresponding true images. (c) Predicted images using the speckle patterns in (b) and the neural network. Correlation values between the true images (a) and the predicted images (c) are presented at the bottom of the predicted images.}
    \label{nature}
\end{figure}

Firstly, we reconstructed hyper-spectral images using a deep learning model based on regression. After training the deep learning model, which is aimed at minimizing the mean squared-loss value between the predicted images and ground-truth images, we reconstruct the letter images using speckle patterns with a 10-by-10 resolution. In Fig. \ref{nature}a, we share ground-truth images for the randomly selected alphabetic letters "AENRTU". These images have a binary phase distribution, as seen here. The corresponding speckle patterns are displayed in Fig. \ref{nature}b. The uniform distribution of bright and dark pixels of the speckle patterns indicates that higher modes of the MMF are excited effectively. With the speckle patterns and the neural network, we reconstructed corresponding letter images, which we call the predicted images as presented in Fig. \ref{nature}c. The predicted images carry high accuracies after comparing them with the ground-truth images. Interestingly the letter image N has an accuracy of 99.9\% even though the letter N shares some common characteristics with other letters such as U, E, and M. Considering the resolution of the speckle patterns (10-by-10) for the training of the deep learning model, these accuracies for image reconstruction are ground-breaking results.

\begin{figure}[!htb]
    \centering
      \includegraphics[width=80 mm]{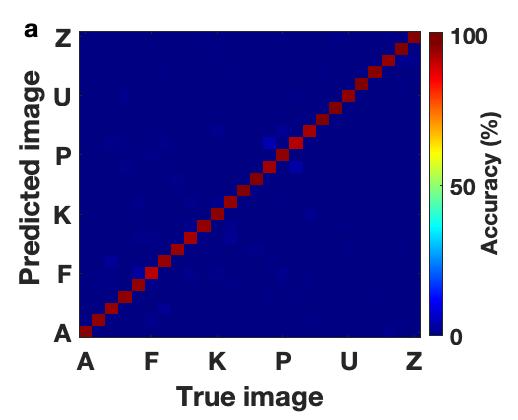}\\
    \includegraphics[width=80 mm]{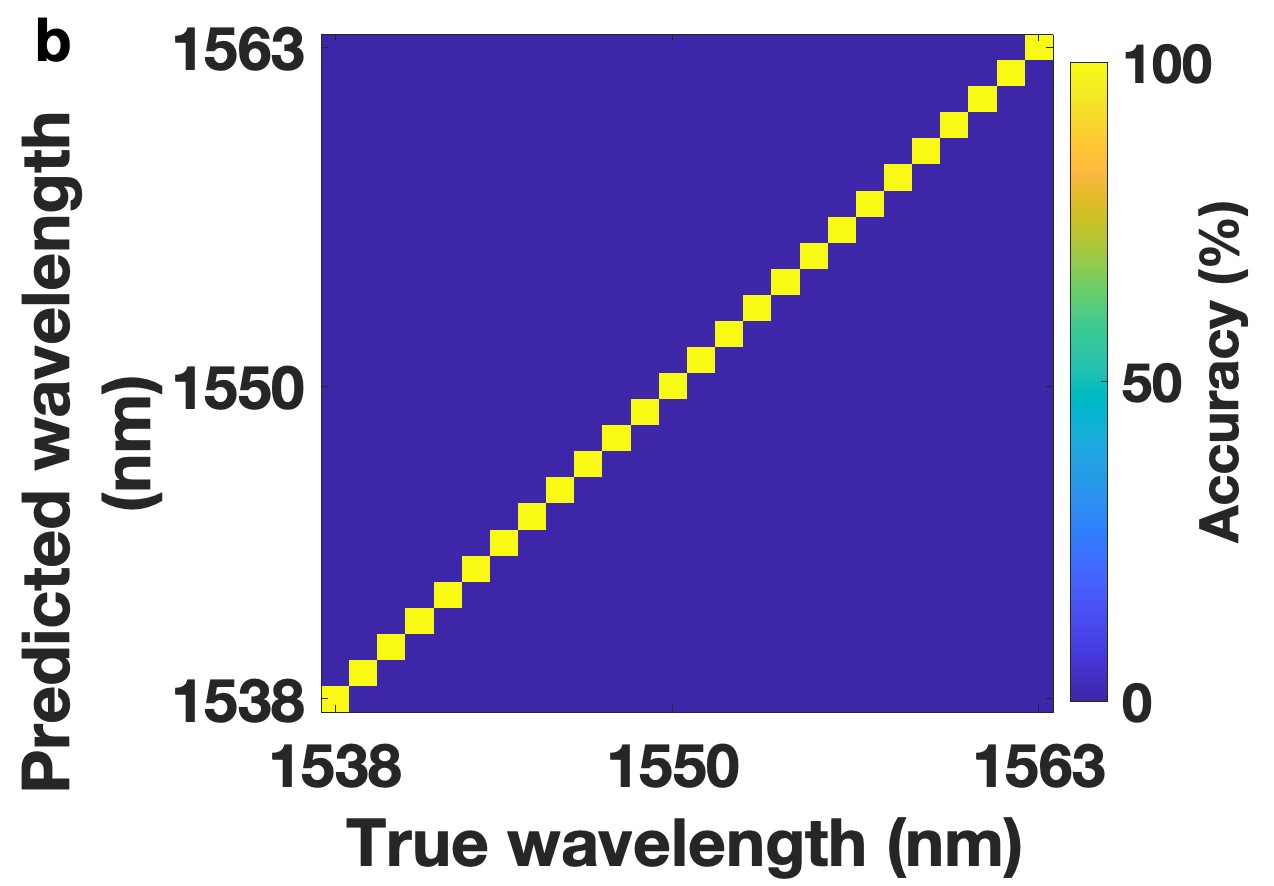}
      
    \caption{Accuracy maps for classification of the speckle patterns in terms of (a) letter and (b) wavelength.}
    \label{classification}
\end{figure}

Later, we classify speckle patterns in terms of the letter after tuning the model's weights. With Fig. \ref{classification}a, we conclude that accuracies of image prediction for many letters are around 95\%. We see that identification of letter F encounters the lowest prediction accuracy as 91.7\%, and high confusion of letter F with letters E and I is seen with 0.7\% accuracy for both letters. We observe that the mean accuracy of image classification is 95.5\% even though speckle patterns for different letters show a high correlation ranging between 99.3\% and 99.9\% (see Fig. \ref{maps}b). Our deep learning model effectively detects the letter images using highly correlated speckle patterns with a 10-by-10 speckle patterns resolution. Then we label speckle patterns in terms of a wavelength channel which ranges between 1538 nm and 1563 nm. The speckle patterns at 1546 nm and 1548 nm are identified with the lowest accuracy, which is 99.5\% for each wavelength channel, and we observe that the mean accuracy of color classification is 99.9\% in Fig. \ref{classification}b. Due to the lower correlations among speckle patterns at different wavelengths of the laser (Fig. \ref{maps}c), speckle patterns for identification of wavelength channels provide us to obtain a greater mean accuracy.

\begin{figure}[!htb]
    \centering
    \includegraphics[width=80 mm]{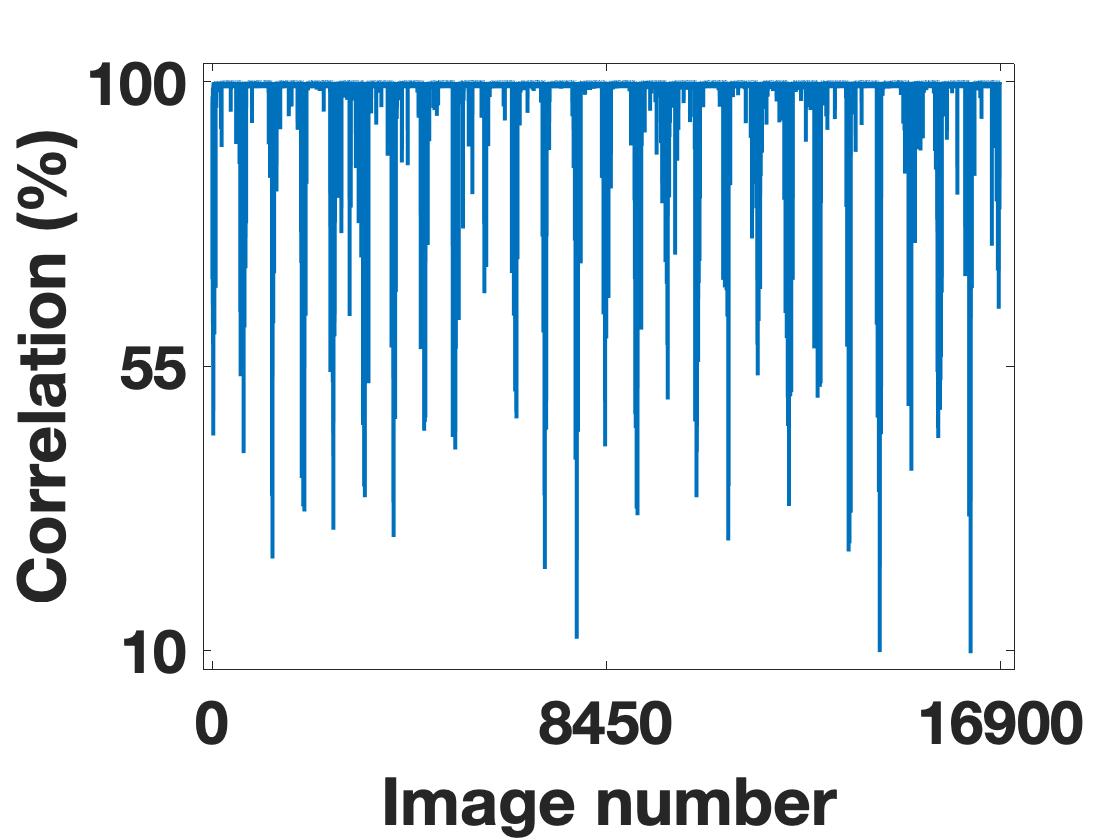}
        
    \caption{The correlation values of the predicted images for all speckle patterns within the data set. }
    \label{imagecorrelation}
\end{figure}

To test the model's performance comprehensively, we reconstruct all letter images using their corresponding speckle patterns within the data set. Then we correlate the true images with the predicted images as seen in Fig. \ref{imagecorrelation}. We encounter high variation in correlation values for the letter images due to high similarities among the speckle patterns within the data set, leading to poor resolving the letter images from the speckle patterns. Fortunately, the mean correlation value is 98\% for 16900 images. This high mean correlation brings superior performance to monitor high-resolution images with the micron size MMF. Here we also see that reconstruction of the images from the speckle patterns requires at least a similar resolution to the speckle patterns utilized for classification of the speckle patterns.

Here, we explore whether neural networks can learn the nonlinear relation between the input images and the speckle patterns. We see that the neural networks infer the speckle patterns using a low resolution of the speckle patterns without any prior knowledge of the light propagation in the MMF, even though the speckle patterns convey high correlations among each other. Our results could be enhanced by using different regions of speckle patterns to correlate the same images, so many speckle patterns will be assigned to the same images while training the deep learning models. Also, increased speckle pattern resolution improves reconstruction and classification accuracies, but unfortunately, the training duration will be longer. Our image propagation approach and deep learning models could transfer images through multiple scattering media such as biological tissues. A transmission matrix of an MMF for imaging requires a ratio of speckle resolution to image resolution is at least two to attain image information successfully\cite{Dremeau2015}. Instead of obtaining the TM, our study demonstrates highly accurate image reconstruction and classification with neural networks and low-resolution speckle patterns.

When we decrease the resolution of the monochrome camera from 256-by-320 to 10-by-10, the frame rate of the camera changes from 344 frames per second (fps) to 10650 fps. This high rise provides more significant information bandwidth. However, the data transfer rate of our work is limited to the wavelength-swept frequency of the tunable laser source (50 Hz) and the refresh rate of the SLM. The refresh rate of the images is boosted by operating a DMD instead of an SLM. The DMDs reach up to a 32 kHz refresh rate, but the DMDs attenuate light power since the DMDs supply amplitude modulation of the light\cite{Conkey2012}. After replacing the tunable laser with a broadband laser and operation of a diffractive optical element\cite{Yolalmaz2021a} to concentrate different wavelengths of the broadband light on different portions of the monochrome camera, multiple speckle patterns for different frequencies of the broadband light are captured simultaneously to speed up hyper-spectral imaging. In this work, we present imaging using a low resolution of the speckle patterns and the neural networks in addition to the robust laser source that provides energy-efficient light transfer and requires a low laser power.

Light propagation through the MMFs is sensitive to tiny external perturbations, leading to the propagation of the input field becoming impractical. With mechanical perturbation on the MMFs\cite{Resisi2020} and transfer learning\cite{Pan2010}, our work will be enhanced for image reconstruction and classification from speckle patterns captured for different geometrical configurations of the MMF. The mechanical perturbation of the MMFs also provides much more degrees of light control than shaping the incident wavefront by the SLMs and DMDs, which results in an increased amount of information propagating through the MMFs. Also, the physics-informed deep learning models may improve the accuracies of our models\cite{Raissi2019}.

\section{Conclusion}

In this study, first time to our knowledge, we show image reconstruction and classification using speckle patterns at the commonly used telecommunication wavelengths. We increase the bandwidth of the imaging system by using a tunable laser propagating through the MMF and keep the number of image pixels the same which is reconstructed by utilizing a lower number of speckle pixels. The speckle patterns are inverted back into the original images on a dynamic diffractive optical element: the SLM. High-accuracy image classification from the speckle patterns using deep learning architecture is seen even though the speckle patterns present high similarities among each other. Furthermore, reconstruction of the images using a small region of the speckle patterns is achieved with a mean correlation of 98\%. Specifically, we show that images belonging to the same class can be reconstructed with up to 99.9\% fidelity without requiring the phase of speckle patterns on the monochrome camera or hyper-spectral transmission matrix of the MMF. We believe that our models are applicable to retrieving significantly complex natural scenes from their speckle patterns.

\section{acknowledgments}

This study is financially supported by The Scientific and Technological Research Council of Turkey (TUBITAK), grant no 118M199, and TÜBA-GEBİP project.

\section*{DATA AVAILABILITY}
The neural network models are available as supplementary material.

\bibliography{Library_Hyperspectral.bib}

\end{document}